\documentclass{article}

\usepackage[british]{babel}
\usepackage[utf8]{inputenc}
\usepackage[T1]{fontenc}
\usepackage{graphicx}
\usepackage{booktabs}
\usepackage{caption}
\usepackage{subcaption}
\usepackage{lmodern}
\usepackage{microtype}
\usepackage{textcomp}
\usepackage{amssymb}
\usepackage{epstopdf}
\usepackage{capt-of}
\usepackage{amsmath}
\usepackage{float} 
\usepackage{xfrac}
\usepackage{dblfloatfix}
\usepackage{url}
\usepackage[squaren,Gray]{SIunits}
\usepackage{commath}

\usepackage{arxiv}

\usepackage[utf8]{inputenc} 
\usepackage[T1]{fontenc}    
\usepackage{hyperref}       
\usepackage{url}            
\usepackage{booktabs}       
\usepackage{amsfonts}       
\usepackage{nicefrac}       
\usepackage{microtype}      
\usepackage{graphicx}
\usepackage{epstopdf, epsfig}
\usepackage{amsmath}
\usepackage{graphicx}
\usepackage{graphics}
\usepackage{listings}

\title{A field investigation of the friction between cross-country skis and snow}
\author{
    Rasmus Nes Tj{\o}rstad \\
  Department of Mathematics\\
  University of Oslo\\
  \textit{rasmusnes.t@gmail.com }
   \And
  Jean Rabault\\
  Norwegian Meteorological Institute \& \\
  Department of Mathematics\\
  University of Oslo\\
  \textit{corresp. author: jean.rblt@gmail.com} \\
   \And
  Olav Gundersen \\
  Department of Mathematics\\
  University of Oslo\\
  \textit{olavgun@math.uio.no}
   \And
  Atle Jensen \\
  Department of Mathematics\\
  University of Oslo\\
  \textit{atlej@math.uio.no}
}

\begin{document}
\maketitle

\begin{abstract}
    Cross country skiing is an interplay between the active motion of the body
    of the skier and the physical interaction between the surface of the skis
    and the snow. Friction and glide between the ski base and the snow depend
    on the snow temperature, ski surface chemistry and roughness, dry friction
    with the snow crystals and liquid meltwater film flow properties. These are
    multi-scale phenomena with physics ranging from the thickness of the water
    film (< 100 nm) to the size of snow crystals ($\sim$ mm) and the length of
    the skis (1-2 m). Although a great amount of resources are being spent on
    understanding glide and friction (the national Norwegian team in skiing
    alone used 15 million NOK in 2014/2018), a basic
    physical understanding of the phenomena at play is still at large and the
    methodology used for assessing the gliding quality of different full scale
    configurations is still coarse in real world applications, which limits
    novel development of robust methods to control and optimize glide. Such
    understanding and full scale testing is particularly important for
    designing new ski gliders, especially in the present context where fluor-based
    formulations are being forbidden because of ecological concerns. In the
    present work, we develop a novel experimental setup and use it to
    investigate through field experiments the basic mechanical phenomena that
    influence glide and friction of full scale skis on snow. The field setup
    features a mobile, novel linear tribometer allowing investigation with
    whole skis in realistic cross-country conditions (speed and load). The
    coefficient of friction between the skis and snow is obtained from a load
    cell, which reports the force necessary to pull a ski sled along a flat
    track. Simultaneously, thermistors measure the temperature at the ski-snow
    interface. In addition, a GPS measures displacement and velocity.
    Coefficients of friction ranging from $\mu_k=0.033$ during cold weather
    ($T=-11^o$C) to $\mu_k=0.0145$ during warm weather ($T=0^o$C) are obtained.
    The highest temperature increase at the ski-snow interface ($\Delta T
    \approx 5^o$C) is measured 40 cm from the ski tip at the highest velocity
    of 8 m/s. The results obtained are in good agreement with several
    measurements reported from similar experiments, and provide a real-world
    validation of results obtained in the laboratory. In addition, we release
    the general tribometer setup as open source material in order to provide a
    foundation on which other research teams as well as cross-country teams can
    further iterate when designing later experiments.
\end{abstract}

\section{Introduction}

Frictional heating and the existence of a partial liquid meltwater layer at the
interface between skis and snow is widely considered to be the main mechanism
causing low friction when skiing. The importance of a lubrication layer has
been known for a long time, as early results \cite{bow39} found that the low
kinetic friction of ice is due to a partial meltwater layer formed by
frictional heating. While it had been previously considered that pressure
melting was alone responsible for the formation of this water film, this early
study found that the coefficient of kinetic friction is independent of the
load, apparent area of contact, and speed of sliding over a certain range, but
increases with decreasing temperature, which suggests frictional heating to be
the dominating effect. Experiments performed on snow showed similar trends -
though the obtained coefficient of friction was higher. The increase in
friction in the case of snow was attributed to the extra work done in
displacing and compressing snow grains. To test this theory further,
\cite{ambach1981ski} developed a probe for measuring the water film between a
gliding ski and snow. The water film was measured to be several micrometers
thick (5-30 $\mu$m), but this is generally considered to be too high, due to
the fact that the measurement device had a hydrophilic coating. In order to
take into account the diversity of mechanisms influencing ski friction on snow,
\cite{colbeck1992review} presented a mathematical model to describe the sliding
process and proposed that the total friction could be expressed as the sum of
several terms representing each mechanism playing a role in the friction
coefficient. This includes contributions from plowing, solid deformation, water
lubrication, capillary attraction and surface contamination. Under most
circumstances of subfreezing conditions, it was found that the heat available
was insufficient to generate enough meltwater to separate the surfaces
completely. Thus it was suggested that both solid-to-solid interaction and
meltwater lubrication play a role in ski-snow friction.

More recently, \cite{strausky1998sliding} attempted to detect possible water
films using fluorescence spectroscopy combined with a pin-on-disc tribometer.
Since the detection limit of the experiment was 0.1 $\mu$m and no film was
detected, it was concluded that water films, if present, must be below this
limit. This value is much smaller than earlier predicted. However, the
experiment was limited to velocities below 0.1 m/s. Following this study,
\cite{buhl2001} examined the gliding of polyethylene on snow in laboratory
experiments and in field tests under racing conditions. The main focus of the
investigation was on the influence of snow temperature and ski surface loading
during gliding. Both laboratory and field experiments showed that kinetic
friction is lowest at around -3$^o$C and increases for low temperatures as well
as for snow temperatures close to 0$^o$C. As the snow temperature increases,
the influence of pressure decreases, and for temperature above $-6^o$C there
was no detectable difference between different loads.

On a more theoretical standpoint, \cite{baurle2006sliding, baurle2007sliding} used a hydrodynamic
approach to explain the measured friction forces and temperature evolution. It
was found that the main factors determining friction are the thickness of the
water film and the relative real contact area. These results also indicate that
the low friction observed is due to unevenly distributed thin water films, with
calculations suggesting that they have a thickness of 30-250 nm along the
slider at -5$^o$C. Moreover, it was found that the behavior of the water films
and size of the real contact area can explain the friction process, and that no
capillary attachment is needed to explain for the experimental results.
Therefore, it is expected that the real contact area is the most critical
parameter determining friction between skis and snow or ice. Following these
findings, \cite{schindelwig2014temperature} studied the temperature of snow
under skating skis with infrared sensors and measured the coefficient of
kinetic friction using a linear tribometer. The highest increase in temperature
(up to 4$^o$C) was measured 10 cm behind the ski binding, in relatively good
agreement with the predictions of \cite{baurle2006sliding}. These results were
further refined in the laboratory \cite{lever2018mechanics} by using
high-resolution infrared thermography (15 $\mu$m resolution) to observe the
warming of stationary snow under a rotating polyethylene slider. No melting at
contacting snow grains was observed despite low friction values being obtained,
which challenges whether meltwater produced by frictional heating is the
dominant mechanism underlying the low kinetic friction on snow. However, due to
the nature of this laboratory experiment, the sliding speed used (0.3$-$1 m/s)
and slider pressure obtained ($0.8-3.6$ kPa) were relatively low compared to
that of cross-country skiing, which raises questions about their applicability
to real skiing conditions.

As a consequence of these limitations in the regimes investigated in laboratory
experiments, more field data are needed to provide an accurate picture of the
mechanisms of importance during cross-country skiing. This is especially relevant in
the current context, where most of the high-performance ski gliders - which are
based on fluoro compounds - are soon going to be banished because of
environmental and health issues \cite{carlson2020ski, skiref}. Therefore, replacements need to
be found, which requires better understanding of the mechanisms at stake and
simpler and more robust protocols for testing skis in realistic concurrence
conditions \cite{kirby2014evaluating, brodie2008fusion, kondo2015estimation}. In this context, developing new
methodology to allow realistic testing of ski surface preparation and gliders
is critical for the performance of athletes in future races \cite{giesbrecht2010polymers, braghin2016engineering, rohm2016effect, budde2017high}, and the aim of
this work is, therefore, to both develop such a methodology and to investigate
the influence of sliding velocity, surface roughness of the ski base, and snow
temperature on the friction during gliding tests in the field, in order to help
compare with laboratory experiments. 

In the following, we start by describing the methodology used for performing
our measurements, including the design of the instrumentation and tribometer used, the
conditions encountered during field testing, and the way we perform our data
analysis. Next, we present our results and compare them with previous
experiments. Finally, we discuss how our experiments could be further
improved and developed into a standardised test procedure.

\section{Methodology}

In this section, we present the experimental setup, including different
components of the mobile
winch system and the instrumentation deployed on the ski sledge,
and the method used for processing the data acquired.

\subsection{Experimental setup and winch design}

The experimental setup was built at the Hydrodynamics Laboratory, University of Oslo. The main components
of the mobile tribometer are a winch and control system which pulls the ski sled,
and an electronics box which sits on the sled and performs measurement and
logging. All the code is released as open source (link: [to be released upon
publication]), and the electronics hardware is based on the open source Arduino microcontroller
boards and other widely available components. This allows to make our setup
reproducible, and more generally using open source designs and code allows to
reduce cost, as was already discussed in several other papers in geosciences
\cite{rabault2016measurements, rabault2017measurements,
rabault2020open}. In the following, we give further details about the design
and functioning of each of these modules.

\begin{figure}[]
    \centering
    \includegraphics[width=0.42\linewidth, height=.3\textheight]{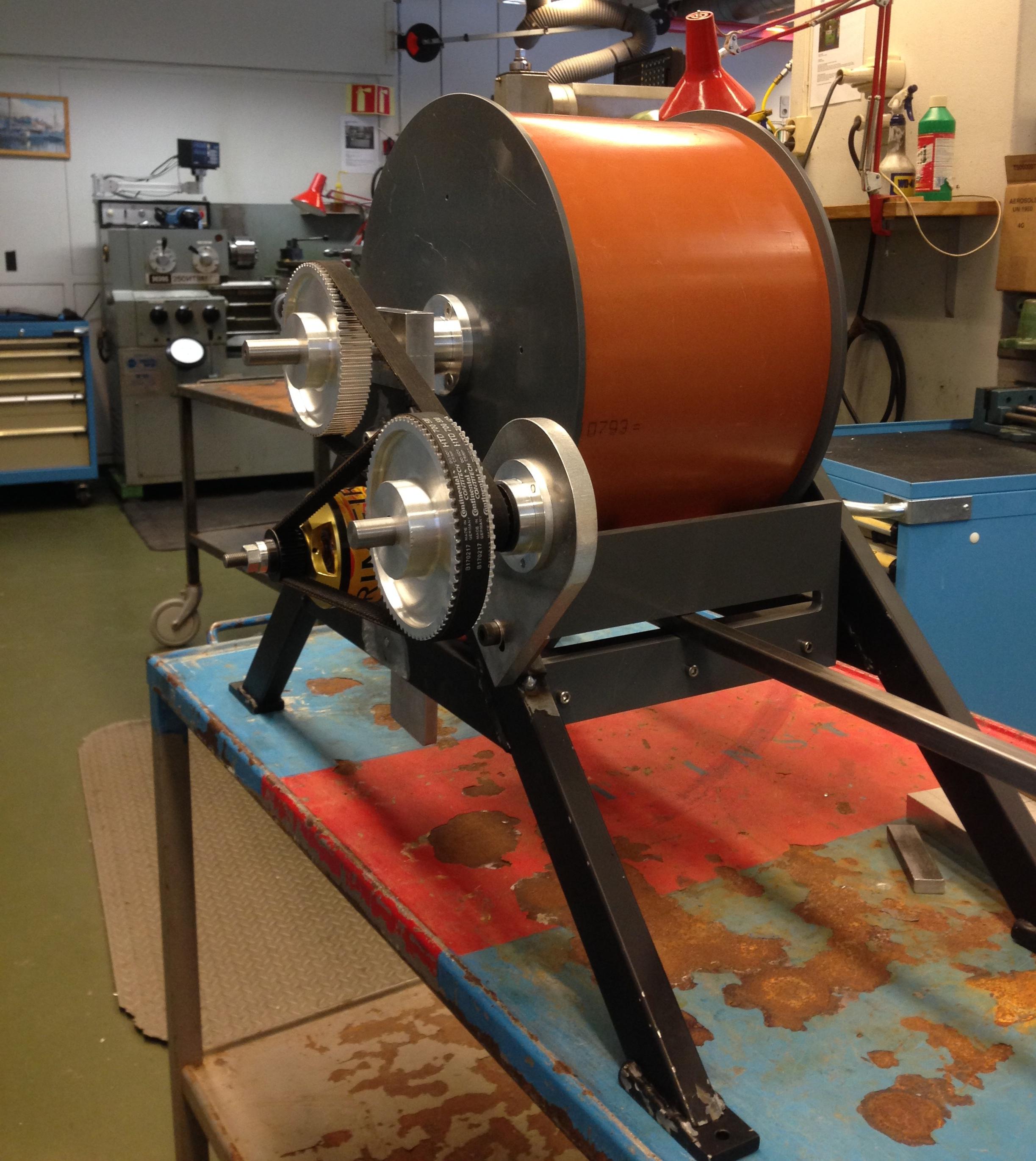}
    \includegraphics[width=0.42\linewidth, height=.3\textheight]{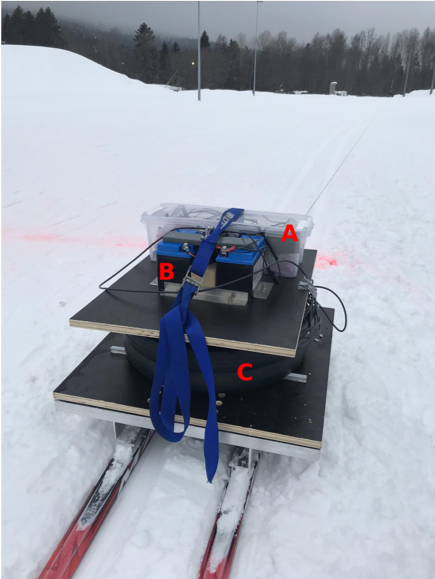}
    \caption{Left: the winch used in the experiment. The drum, steering
    mechanism, motor and two stage gears are clearly visible. Right: the sled on snow,
    ready for a run. The box containing the electronics (A), motorcycle
    batteries (B) and added weights (C) are clearly visible.}\label{vinsj1}
\end{figure}

The winch consists of a custom-built winch drum (diameter 31.5 cm) mounted on
support legs, as visible in Fig. \ref{vinsj1}. The setup is mobile, but it is
important that the construction is solid and heavy enough to withstand the
motion of the drum when pulling the sled. Indeed it is crucial that the winch
is stable to ensure that the measurements obtained are due to motion of
the skis and not wobbling from the winch. Therefore, three L-shaped aluminum
profiles were attached to the bottom of the frame on which the winch is
mounted. These profiles reach 20-25~mm into the snow and hold the winch in
place. In addition, three 12V lead acid car batteries used for powering the winch were placed
on the same frame during the runs, further securing that the winch does not
move during the experiments. The winch has a steering mechanism for the rope
which allows to guide the rope while reeling in the sled, thus enabling a
uniform distribution of the rope over the drum. The rope used in the experiment
is a 150 m long static climbing nylon rope with minimum breaking strength of 100 kg-force
and a typical elasticity of 2\%.

The motor used in the setup is a Rimfire 65CC (90-85-160) (Great Planes Model Manufacturing, 
Champaign, Illinois, United States) brushless electric
motor with a maximum continuous power output of 7500W. It is driven through a
Phoenix Edge Lite HV160 (Castle Creations, Kansas 66062 USA) motor controller, which can accept up to 50~V input and
is rated to deliver up to 160~A, well over the peak power consumption we
expect and the 36V power supplied by the 3 car batteries. Both elements are
available from flying models hobby shops,
which also means that these are cost-effective alternatives to more specialized
industrial solutions. The power output of the motor is set by an open source
Arduino microcontroller (Arduino Uno, Arduino foundation, Turin, Italy), which is connected to the interface of the motor
controller and emulates a radiocommand output for communication and control. The Arduino is also equipped
with a Long RAnge (LoRA) transmission module (Adafruit Feather 32u4 RFM95 LoRa Radio, Adafruit Industries, New York, New York, USA), which allows two ways
communications with the electronics box on the skis sled. The use of an Arduino
allows to run predefined motor power profiles, and to perform reproducible test
runs. In the present case, fixed motor power output was considered as a
satisfactorily way to control the profile of each run, but a rotary encoder
could easily be added to the drum wheel to perform closed-loop control if found necessary. The motor drives the drum through a two stage gear
system, giving a reduction factor for the rpm of around 13:1 (two stages of
both 3.6:1 are used successively for the rpm reduction). Power is provided by
the three car batteries mounted in series (total voltage 36V), which are also
used for stabilization of the frame. Finally, a high current spark-safe switch
is added, which allows to perform a hardware cutoff of the engine if necessary.
The Arduino board is connected to a computer that is used for providing control
commands to the whole system.

The sled is composed of a U-frame on which two cross country skiing
fixations are mounted. The distance between the fixations can be adjusted to
perform experiments on different ski tracks. The axial direction of the
fixations is fixed, in order to avoid parasitic motion of the frame and provide
rigidity to the skis-sled assembly. A plate is fixed onto the frame, and
equipped with a vertical bar on which training weights can be secured. A second
plate is mounted above the weights, which receives the electronics used for
measuring and logging the physical quantities of interest as well as the two
motorcycle batteries assembled in series (total voltage 24V) used to power the
sensors, and the electronics through a low dropout voltage regulator. Skis can
be easily and quickly swapped using the fixations, which allows to test different grinds,
base gliders and topping wax within a short time period. A picture of the assembled sled is
visible in Fig. \ref{vinsj1}. By varying the amount of weights used, the total
mass of the assembly can be varied between around 13 and 100~kg.

We had two pairs of skis at our disposal, one pair with cold grind and the
other one with wet grind. Both pairs were provided and prepared by the Norwegian Olympic
team (Olympiatoppen), who also provided the details of the grinds properties.
The cold grind has the finest roughness ($R_a \approx
2.5 \mu$m) and is usually used under cold and dry conditions. Its roughness
yields a larger area of contact that should result in higher friction and
frictional heat at the front part of the skis and therefore help produce
meltwater in cold conditions. The wet grind in contrast has a more coarse
structure ($R_a \approx 5 \mu$m) and is used under warmer and wetter conditions, when
one wants to reduce the contact area with the snow in order to avoid excessive
production of water at the ski-snow interface.

Two kinds of sensors are mounted on the skis and sled. First, a S2M load cell
manufactured by HBM (HBM Gmbh, Darmstadt, Germany) is mounted at the front of the sled. It is used to measure
the pulling force by the rope on the sled. The rope is attached to the cell
through two knuckle eyes, which prevents the application of bending moments.
The load cell has a typical sensing range of 200~N, with a sensitivity of
2~mV/N. It also incorporates an overload stop force limit, which protects the
cell if higher transient forces are applied. The load cell signal is amplified
before being logged by the Arduino board (see the Electronics section for more
details).

\begin{figure}
    \centering
    \includegraphics[width=0.45\linewidth]{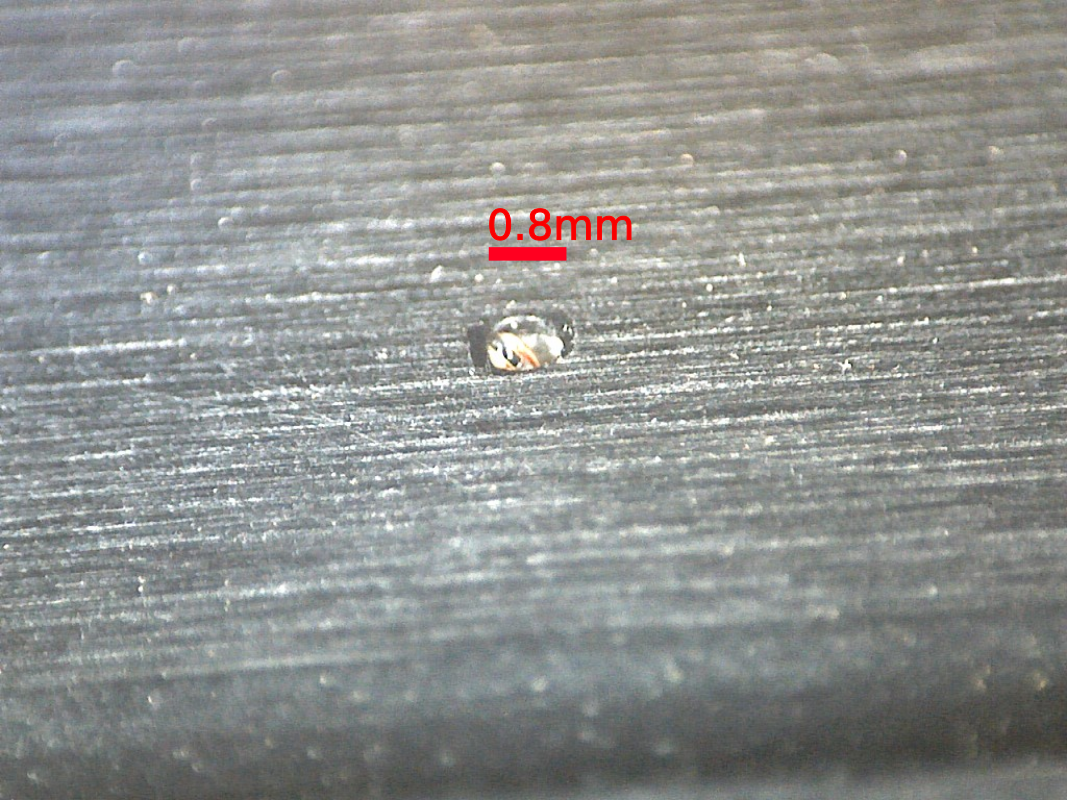}
    \caption{A thermistor in the ski viewed through a microscope. The picture is taken with an angle of 45 degrees.
    The sensor is flush mounted with respect to the base of the ski. The diameter of the sensor
    head is 0.8mm}\label{thermistor_in_ski}
\end{figure}

Second, an array of NTC thermistors (B57540G1, TDK Corporation, Tokoyo, Japan) is mounted on one of the two
skis, at the ski-snow interface. Using thermistors allows to perform
temperature measurements with both high accuracy and frequency, at the cost of
a requirement for calibration before the beginning of each series of runs. The
thermistors are set in place through holes in the skis, and adjusted by hand so
that the tip of the glass enclosure is level with the base of the ski,
therefore measuring the temperature at the interface with the snow. Being very careful
adjusting the height of the thermistors is critical for reliable temperature measurements,
as in our experience, having the thermistors too long outside of the skis means that
the temperature of the snow rather than the snow-ski interface is reported, while having
the thermistors not lowered enough means that the temperature of the plastic core of
the ski is measured. Therefore, the position of the thermistors where carefully done, and was controlled under a microscope, 
as depicted in Fig. \ref{thermistor_in_ski}.

\begin{figure}
    \centering
    \includegraphics[width=0.45\linewidth]{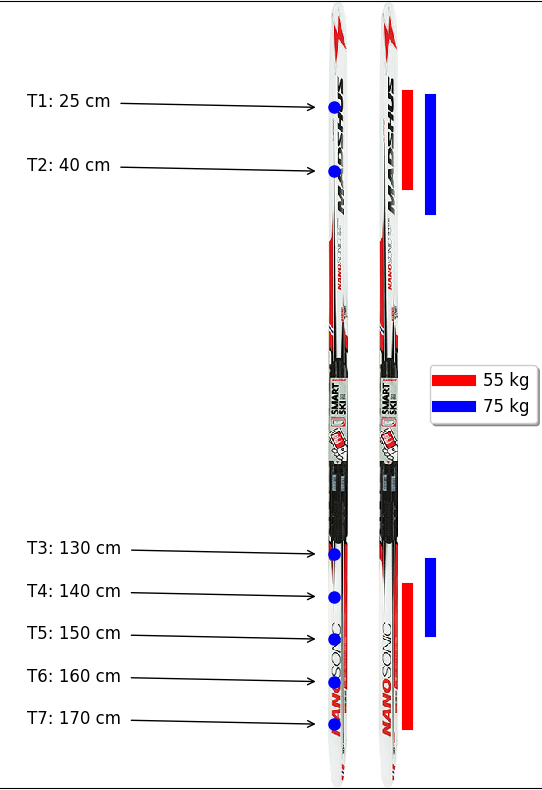}
    \caption{The position of the thermistors (T1 to T7) along the ski. As the
    skis bend under the weight of the skier, and have a natural curvature at
    rest, the region of the skis in contact with the snow depends on the weight
    applied on the skis (lateral bars indicating the friction region for a
    loading of 55 and 75kg, respectively).}\label{positionThermistors}
\end{figure}

A socket
connector is used to connect the thermistors to the electronics box, which
allows to quickly swap skis. The position of the thermistors is visible on Fig.
\ref{positionThermistors}, and covers both the front and rear part of the ski
that are in contact with the snow. As the skis bend under the weight of the
skier, and have a natural curvature at rest, the region of the skis in contact
with the snow depends on the weight applied on the skis (lateral bars indicating
the friction region for a loading of 55 and 75kg, respectively). The kicking
motion of the skier when going forward explains why it is relevant to study the
full area in contact with snow when a large load (75 kg) is applied. Depending
on the load applied, relevant different thermistors of interest will be
presented in the following.

The system-level overview of the winch setup is visible in Fig. \ref{fig:experimental_setup_schematic}.

\subsection{Electronics on the ski sledge and data acquisition}

The electronics box mounted on the sled is composed of an Arduino
microcontroller, a LoRA module for communications with the control system of
the winch (analogous to the one on the winch),
a SD card module for storing data, a GPS and antenna, a series of
Analog to Digital (ADC) shields for logging the thermistors, and an amplifier
(AE301 from HBM, minimum non-linearity, relative reversibility error and lateral force effect of
0.02 \% each) to convert the force gauge measurements to a range easily
measurable by the 10 bit ADC built in the Arduino. The sampling frequency of
the force gauge output is set to 1kHz, and the amplifier is set up so that the
10 bit resolution is obtained for a load between 0 and 20 kg-force. This results in
a theoretical resolution of 0.2N per ADC step.

Thermistors reflect changes in the temperature of the semiconducting material. To measure this
change in resistance, they are assembled in series with a known resistor. This
constitutes a voltage divider, and by applying a fixed voltage drop over the
assembly, the resistance of the thermistor can be calculated from measuring its
voltage drop. More specifically, if $V_0$ is the imposed voltage drop over the
thermistor and resistor assembly, $R_0$ the known value of the resistor, $V_T$
the voltage drop around the thermistor, and $R_T$ its resistance, one gets
that:

\begin{equation}
  R_T = R_0 \frac{V_T}{V_0 - V_T}.
\end{equation}

We resort to ADC shields to log the voltage drop across the thermistors in
order to increase the resolution of our measurements. The shields chosen (ARD-LTC2499 ADC, 
Iowa Scaled Engineering, Elbert, CO 80106 USA) include a 24-bit ADC and can be stacked on top of each
other. While this ADC offers a good resolution of 24 bits, it is quite slow
(the maximum sampling rate in free mode is 15Hz, which is far slower than the
response time of the thermistors), which is the reason why we use one shield
per thermistor and stack the shields rather than further reducing the sampling frequency through a
multiplexing approach. The voltage $V_0$ is set to 4.096V in our case, and it
is provided from a high accuracy power regulator present on the ADC boards to
avoid fluctuations due to temperature or the voltage change of the battery.

The GPS (Copernicus II GPS, Trimble Navigation Limited, Sunnyvale, CA 94085, USA) is able to
measure velocity with an accuracy of 0.06m/s using doppler shift, according to its data sheet. It is highly sensitive (-160 dBm tracking sensitivity and -148 dBm acquisition sensitivity) and supports Satellite-based Augmentation Systems (SBAS) for increased accuracy. Measurements are performed at a frequency of 1Hz, and it is used to estimate the velocity of the sled.

In our experience, the two motorcycle batteries were enough to provide energy
to the whole system for a complete day of experiments. 

\begin{figure}
    \centering
    \includegraphics[width=0.65\linewidth]{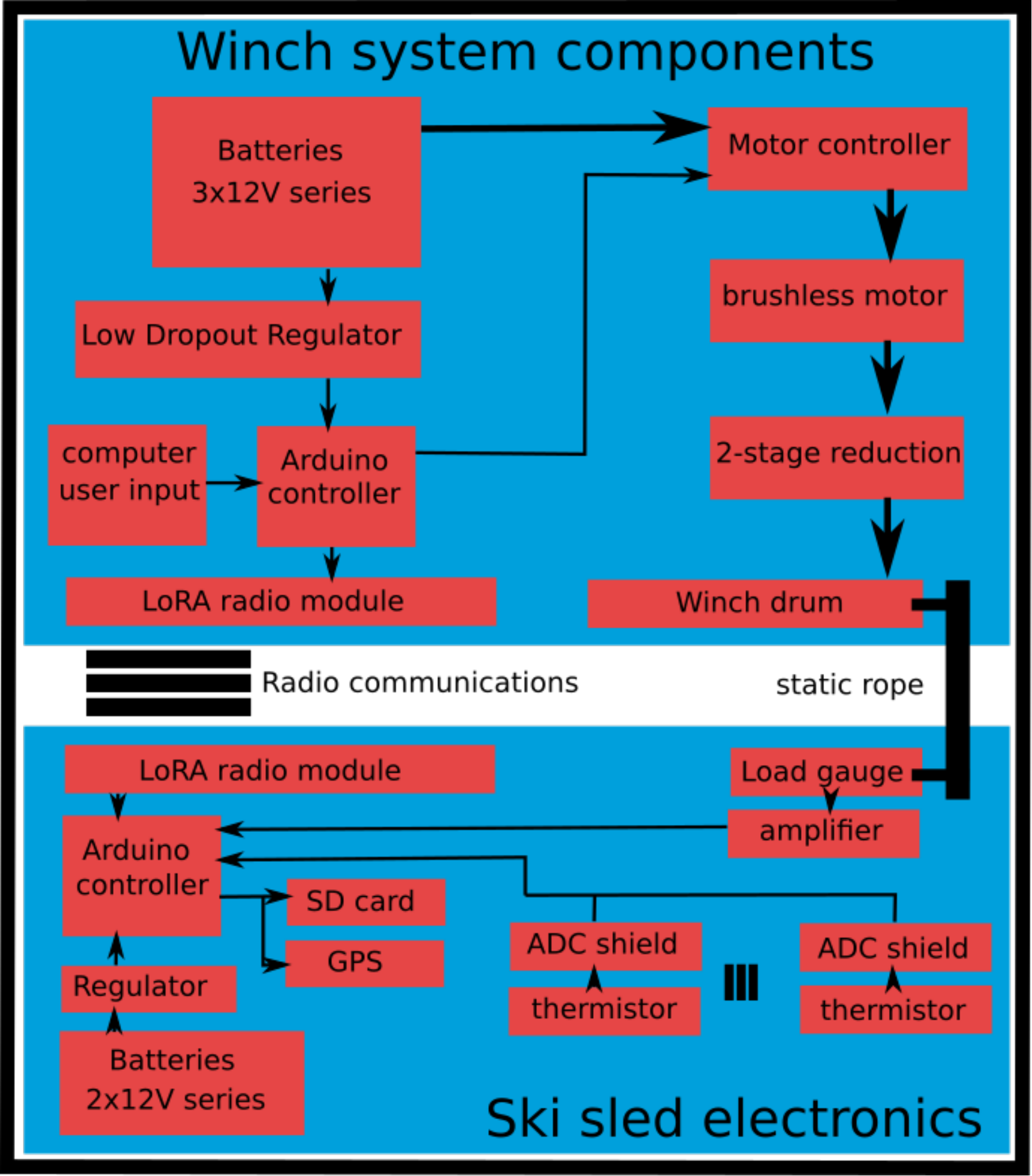}
    \caption{System-level overview of the winch and ski sled, and the communications between the two. Using open source electronics and libraries is a critical element for allowing all sub-systems to communicate reliably with each other, and maintaining affordable costs.}\label{fig:experimental_setup_schematic}
\end{figure}

The system-level overview of the ski-sled setup is visible in Fig.
\ref{fig:experimental_setup_schematic}.

\subsection{Experimental procedure}

When a run should be performed, the skis sled is first put in position away
from the winch and connected to the traction cable through the load cell. Then,
the electronics from both the winch control and the electronics box on the sled
are powered on. At this point, the equipment is ready to perform a run.
Starting the winch traction profile program sends a radio signal to the
electronics box on the skis sled that it should start recording data. The
traction profile is composed of three segments: first, the traction speed
(controlled by the power of the engine) is linearly increased to smoothly
accelerate the sled. Then, a segment of constant traction speed takes place.
The data from this segment is used for data processing. Finally, the
traction is stopped by linearly going to zero. When the run is finished,
another radio signal is sent by the winch control system to ask for end of data
recording, and the incrementing of the logfile index in the logger.

In the present experiments, velocity is varied between $v=1$ and $9$ m/s
by regulating the power output of the motor. The velocity of the sled is
measured using the GPS module. At velocities $v < 1$ m/s, the torque of the
motor is insufficient to surmount the friction. The upper velocity limit
is set by the acceleration phase and the stability of the sled. To get as
accurate results as possible for the coefficient of friction, it is necessary
that the sled has a constant velocity for as long as possible. For higher
velocities more of the track will be spent on acceleration and deceleration,
and ultimately will result in no part with constant velocity. We experimented
with a higher acceleration to resolve this issue, but this led to a new problem
due to the elasticity of the rope. Indeed, the rope is supposed to have very
little elasticity and to be inextensible, but over a distance of 100-150 meter
this was not strictly the case. Over this distance it was possible to stretch the
rope up to 1 to 2 meters. A too abrupt acceleration resulted in a high tension
being built up in the rope before it catapulted the sled down the track. The
sled then achieved a velocity higher than the rate at which the winch pulled the
rope, resulting in the sled sliding over the rope. This led to high noise in
the load cell data and unreliable thermistor data.

A total of seven field experiments were performed between February 28 and April
10, 2018. Details regarding the weather and snow conditions of each test
session are provided in Table \ref{weathertable}. The ambient temperature and
snow temperature during an experiment were measured by a Doric Trend indicator
412A (Doric Instruments, USA, resolution $0.1^o$C, accuracy $0.6^o$C). The snow temperature was
measured approximately 5 cm below the surface and the ambient temperature 30 cm
above the surface. Both measurements were performed in the shade.

The experiments were conducted in cross country tracks located at Sognsvann
Sn{\o}park and Holmenkollen Skiarena, Oslo. In both places a straight part of the track
was used for performing the measurements. The GPS was used to measure the
elevation profile of the tracks. The obtained data were promising, but the resolution is limited to 1 m. The GPS measured a constant value across the entire length of the track in both cases, $h_S = 181$ m and $h_H = 330$ m for Sognsvann Snøpark and Holmenkollen, respectively. The track section used at Sognsvann is approximately 120 meters long. The most critical downside with the track there was the uneven elevation profile which was clearly visible. This made it difficult to find an appropriate interval to use as a measurement section for the coefficient of friction. The first 6 out of 7 field experiments were conducted at this location. The track used at
Holmenkollen was more straight and presented also less variations in elevation,
as confirmed by carthographic data.
A section of approximately 80 meters behind the Biathlon Stadium was used as
test track. Ideally, most of the testing should have been done there, but only
one field experiment was performed at this location since it was hard to find
available time slots due to the many ski events held there (including FIS and
IBU World Cup races).

\begin{table*}[]
    \resizebox{\textwidth}{!}{%
\begin{tabular}{llllll}
\hline
    Date & Time & Ambient temp. ($^o$C) & Snow temp. ($^o$C) & Wind (m/s)& Weather\\
28.02 & 14:00-16:15 & $-11$ & not measured & 4-5 m/s &  Clouded\\
07.03 & 11:30-13:00 & $-2$ &$-4$ & 2 m/s &  Clouded, some light snow fall\\
14.03 & 10:00-14:00 & 0 to 4 & $-6$ to $-4$ & 1 m/s &  Sun \\
15.03 & 13:00-15:00 & $-2$ to $-1$ & $-4$ to $-3$ & 2-3 m/s &  Clouded/sun \\
16.03 & 12:00-14:30 & $-4$ to $-3$ & $-8.8$ to $-7.3$ & 1-2 m/s &  Sun \\
27.03 & 09:00-13:00 & $-4$ to 1 & $-7.6$ to $-6.2$ & 1 m/s &  Sun \\
10.04 &  09:00-13:00 & $-1$ to 5 & $-1.2$ to $-0.1$ & 2-3 m/s &  Sun\\
\hline
\end{tabular}%
}
    \caption{Weather and snow conditions for each of the seven test
    sessions.}\label{weathertable}
\end{table*}

\subsection{Data analysis}

The physical properties to be measured in our experiment are the velocity of
the sled, its friction coefficient, and the temperature at the ski-snow
interface. As previously mentioned, the velocity of the sled is simply obtained
from the GPS data at a frequency of 1Hz.

\subsubsection{Coefficient of friction}
\begin{figure}
    \centering
    \includegraphics[width=0.45\linewidth]{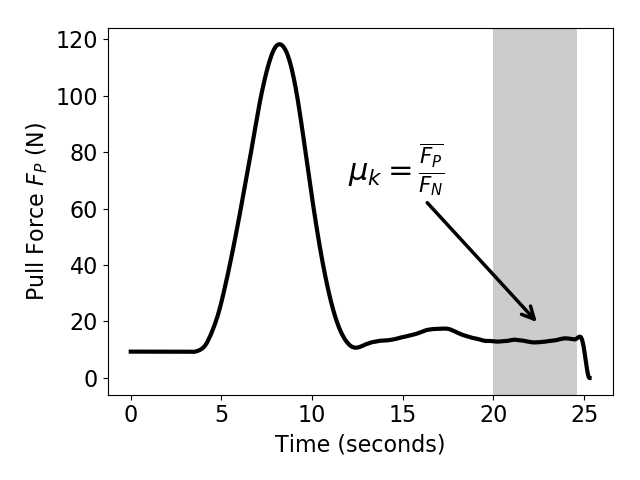}
    \includegraphics[width=0.45\linewidth]{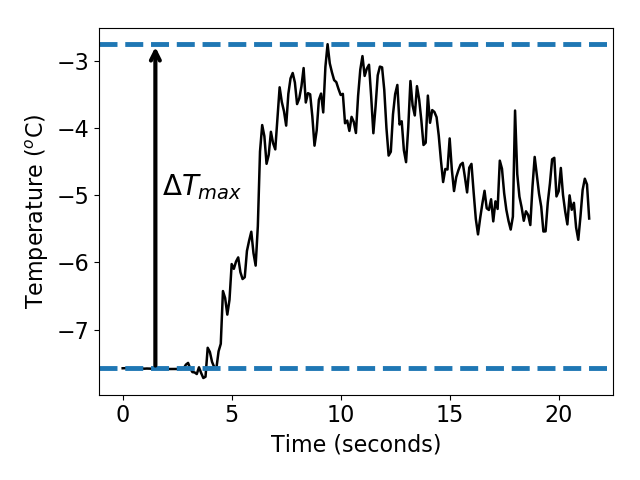}
    \caption{Illustration of the processing applied to raw data of the pull force
    and temperature (the two plots are not from the same run). Left: pull force measured during a representative run. The shaded area
    indicates the measurement section that is used to estimate the friction coefficient ($\overline{F_P} = 13.1$ N in this particular case).
    Right: Representative example of the temperature recorded by a thermistor
    during a run. The sled starts from rest. $\Delta T_{max}$ indicates the
    maximum increase in temperature, which is the quantity reported in the following
    of this work.}\label{fig:both_friction_temperature}
\end{figure}

The coefficient of friction is calculated based on the measurements provided by
the load gauge. The pull force, $F_P$, is a combination of the frictional
forces between the skis and the snow $F_F$, the aerodynamic drag of the sled
$F_A$, the acceleration of the sled, and the slope of the section. The
aerodynamic drag is calculated from the cross-sectional area $A$ of the sled,
the sled velocity $v$, and a typical drag coefficient $C_D$ following $F_A =
1/2 \rho v^2 A C_D$. In our case, $A$ is smaller than what is obtained for a
skier (around 0.09 m$^2$ vs. typically around 0.54 m$^2$ in the case of a
skier \cite{leino1983methods, ainegren2018drag}). For typical friction coefficients ($C_D
\approx 1$) and velocities ($v \approx 2.2$m/s and $ v \approx 3.9$ m/s) used in our experiments, the
aerodynamic drag is less than around 7 \% of the friction force and will
therefore be neglected in the analysis. Hence, in sections of constant velocity
when no sled acceleration is present and the track is horizontal, the friction
force $F_F$ will be regarded as equal to the pull force $F_P$. Due to the
acceleration phase, elasticity in the rope and elevation profile, only a
small part of each test run can be used as a measurement section, as shown in
Fig. \ref{fig:both_friction_temperature}. The coefficient of friction $\mu_k$ is
determined from the averaged force measured in this section and the normal load
of the sled, $<F_N>$, following:

\begin{equation}
    \mu_k = \frac{<F_P>}{F_N}.
\end{equation}

In our reported results, each coefficient value is based on a sequence of runs, where each sequence consists of
approximately 5 consecutive runs with the same grind and velocity, at the same
location. Performing several runs is used as a means to reduce the variability
inherent to field measurements. A sequence of runs could occasionally have
fewer or more runs due to time constraints and weather conditions, and this is
taken care of in the averaging process.

\subsubsection{Temperature}
The temperature at the ski-snow interface is measured by the thermistors
mounted at the ski base. As stated previously, an increase in temperature is
expected due to the frictional heating caused by sliding. In order to obtain
the value of the temperature from the resistance measured across a thermistor,
the Steinhart-Hart \cite{steinhart1968calibration} is used to describe the temperature-resistance
curve following:

\begin{equation}\label{steinhart}
\frac{1}{T} = A + B\ln(R_T)+ C\big(\ln(R_T)\big)^3,
\end{equation}

\noindent where $T$ is the temperature (in Kelvins), $R_T$ is the measured
thermistor resistance (in Ohm), and $A$, $B$ and $C$ are the Steinhart-Hart
coefficients. This is a much used third-order approximation which accounts for the highly
nonlinear relationship between resistance and temperature. The Steinhart-Hart coefficients are obtained from the datasheet of the thermistors, and
fine-tuned through calibration. For fine-tuned calibration, the formula Eqn.
(\ref{steinhart}) was fitted to experimental data obtained for reference
temperatures of -10, 0 and 5 degrees Celsius.

The relevant data that will be presented further in the article are the maximum
increase in temperature $\Delta T_{max}$, and the temperature profile along the
ski. $\Delta T_{max}$ is defined as the maximum difference between the
temperature $T_0$ before sliding and the temperature $T(t)$ observed as a
function of time during the run, as indicated in Fig. \ref{fig:both_friction_temperature}. It
is therefore expected that $\Delta T_{max}$ measures the friction-induced
heating of the skis during gliding.

\subsection{Ski track characteristics}
Due to the lack of equipment for snow
and humidity measurements, the track conditions are described with simple
keywords in order to provide some background information beyond that of
temperature. More specifically, we use the following keywords:
\begin{itemize}
    \item Prepared - The track has been prepared the same day and has
        experienced few passings before the field experiment was conducted.
    \item Glazed/Icy - The track has not been prepped the last 24 hours.
        Furthermore, the track has frozen overnight which results in a hard,
        icy surface.
    \item Transformed - The track has experienced ambient temperatures above
        $0^o$C, triggering changes in the micro-structure of the snow / ice.
\end{itemize}

\section{Results}

In this section, we report the data acquired following the methodology previously
described. Firstly, we analyze the influence of temperature, surface roughness,
sliding velocity and normal force on the coefficient of friction at the
ski-snow interface. Next, we look at how these parameters influence the
temperature evolution at the ski-snow interface, as measured by the
thermistors, and we discuss how these temperature changes may be related to the
skis preparation and the observed friction coefficient.

\subsection{Measurements of friction coefficient}

The averaged friction coefficient is reported in Table \ref{coefficienttable}
along with date, temperature, velocity and snow conditions. 

In the following subsections, the effect of ambient temperature, surface
roughness, velocity and normal load on the observed friction coefficient are
presented individually.

\begin{table*}[t]
    \begin{center}
    \begin{tabular}{lllllr}
    \hline
    Date & Temp. (C) & Snow type & Velocity (km/h)&  \multicolumn{2}{c}{Friction coefficient} \\
    \cline{5-6}
     &  &  &  & Cold Grind & Wet Grind \\
    \hline
    28-Feb      & -11   & Prepared & 8      & 0.0326  &  -    \\
    07-Mar      & -2    & Prepared & 8       & 0.0229 & - \\
    07-Mar      & -2    & Prepared & 14      & 0.0265 & - \\
    14-Mar      & +0    & Prepared  & 8       & 0.0182 & - \\
    14-Mar      & +1     & Prepared and Transformed & 14      & 0.0217 & - \\
    14-Mar      & +2     & Prepared and Transformed & 8       & -     &  0.0182 \\
    14-Mar      & +3     & Prepared and Transformed & 14      & -     &  0.0262 \\
    15-Mar      & -2    & Prepared & 8       & 0.0251 & - \\
    16-Mar      & -4    & Prepared & 8       & 0.0241 & - \\
    27-Mar      & -3    & Glazed/Icy & 8       & 0.0214 & - \\
    10-Apr      & +0     & Prepared and Transformed & 14      & 0.0145 & - \\
    10-Apr      & +3     & Prepared and Transformed & 14      & 0.0179 & 0.0171 \\
    10-Apr      & +5     & Prepared and Transformed & 14      & - & 0.0209 \\
    \hline
\end{tabular}%
    \caption{Table of snow conditions, towing velocity, and obtained friction
    coefficients. All results are averages over several runs performed in a
    short amount of time, in order to reduce scatter and variance of the
    estimates for the friction coefficient.}\label{coefficienttable}
        \end{center}
\end{table*}

\subsubsection{Effect of ambient temperature} 
\begin{figure}[htbp]
    \centering
    \includegraphics[width=0.45\linewidth]{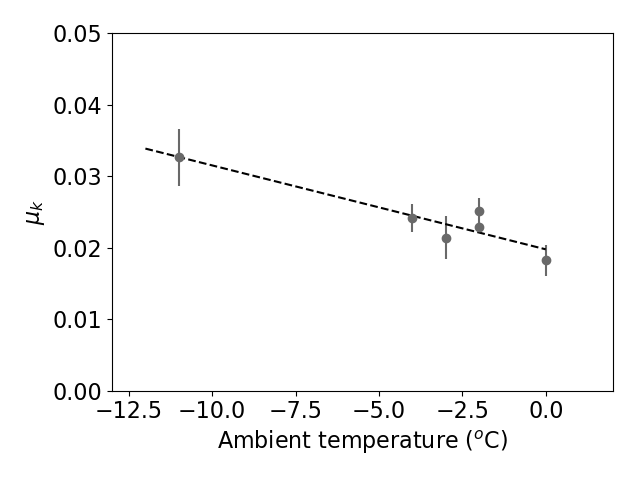}
    \includegraphics[width=0.45\linewidth]{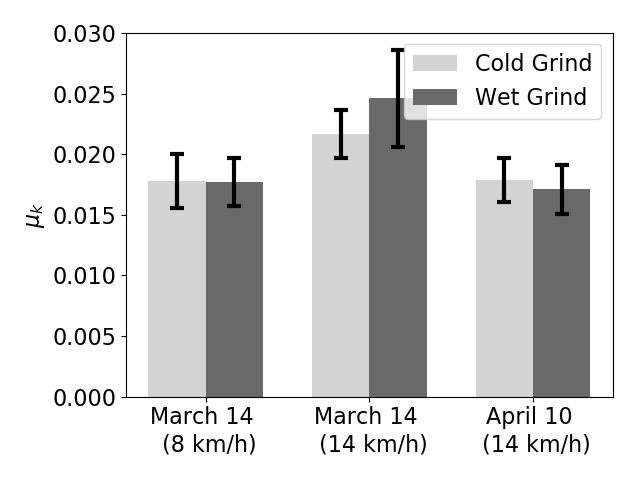}
    \caption{Illustration of the dependence of the friction coefficient on
    experimental conditions. Left: friction coefficient $\mu_k$ for cold grind at constant velocity $v=8$ km/h
    as a function of temperature. The temperatures presented are air
    temperatures measured at the experiment site. The friction coefficient is
    presented with $2\sigma$ confidence intervals. The general trend, i.e. a
    decrease of friction as temperature increases up to the freezing
    temperature of water as indicated by the dashed line, is in agreement with
    previous studies. Right: Mean and error bars (2 $\sigma$) of the friction coefficient
    $\mu_k$ for wet and cold grind at temperatures close to 0
    degrees.}\label{fig:friction_tmp_roughness}
\end{figure}

The dependence of the friction coefficient on temperature has been
confirmed by many researchers. It is a widely held view that the coefficient of
friction decreases with increasing temperature due to enhanced lubrication at
higher temperatures. Fig. \ref{fig:friction_tmp_roughness} shows the measured
friction coefficient $\mu_k$ for six different temperature. The behavior is in accordance with
previously published results, and indicates a decrease of friction with
increasing temperature up until around 0 degrees \cite{bow39, klein1947snow,
ermakov1984coefficient, buhl2001, baurle2006sliding, nachbauer2016friction}.

\subsubsection{Effect of the surface roughness of the skis obtained through a cold grind versus wet grind preparation}

It is a commonly held view that a rougher sliding surface should be used during warm
and wet conditions and that a smoother surface should be used during cold and
dry conditions. \cite{shimbo1971friction} conducted experiments where he found
that friction increases with roughness at sub-freezing temperatures and that
friction decreased with roughness for $T=3^o$C.  Our
experiments were conducted in ambient temperature $0^o$C $ \leq T_{amb}$, hence
we would expect the wet grind to yield lower $\mu_k$ in all cases. The results
are presented in Fig. \ref{fig:friction_tmp_roughness}. In our experiments, by contrast
with the expectations from previous results published in the literature, we
find that the differences in friction coefficient between the wet and the cold
grind were not statistically significant w.r.t. $2 \sigma$.

\subsubsection{Effect of Velocity}

The friction for objects sliding on snow or ice is dependent on velocity.
\cite{oksanen1982mechanism} conducted experiments with sliding friction on ice
and showed that the coefficient of friction increases with velocity for snow
temperatures near zero ($T_{snow} > -1^o$C, $0.5 \leq v \leq 3$ m/s ). For
colder snow temperatures ($T_{snow} = -15^o$C), the friction decreased with
increasing velocity ($0.5 \leq v \leq 3$ m/s). These results can be physically
explained by the presence of a water film. At temperatures near zero, the water
film may already be present, and further frictional heating will generate too
much water and induce capillary drag \cite{giesbrecht2010polymers}. At colder temperatures the frictional
regime is dominated by solid-to-solid contact and the water film generated by
frictional heating will reduce friction. Due to lack of valid data from experiments conducted during cold temperatures, we are only be able to
investigate the friction coefficient dependence on velocity in the warmer
temperature regime.

\begin{figure}[htbp]
    \centering
    \includegraphics[width=0.45\linewidth]{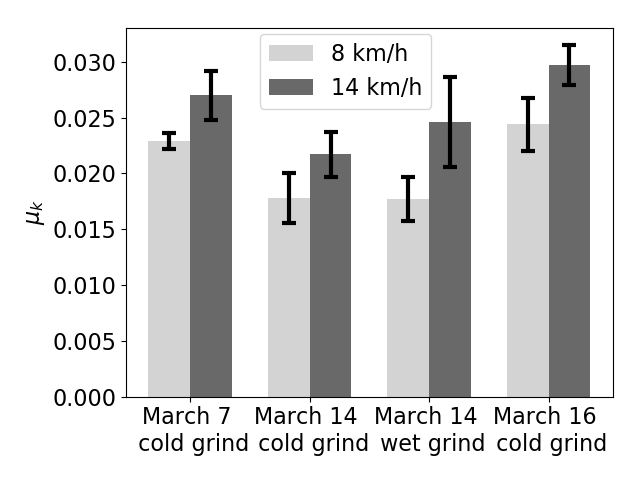}
    \caption{Mean and error bars (2 $\sigma$) of the friction coefficient
    $\mu_k$ for two velocities in different conditions. Three of the datasets
    are from cold grinded skis and one is from wet grinded skis. A clear trend
    of increased friction coefficient at higher sliding velocity is
    visible, similar to what has been reported by other studies in the
    literature.}\label{field_velocity_and_friction}
\end{figure}

As can be seen in Fig. \ref{field_velocity_and_friction}, the friction is lower
for the lowest velocity in all four cases. This result is similar to what was
reported by previous works \cite{kuroiwa1977kinetic, jones1994friction,
baurle2006sliding}.

\subsubsection{Effect of Normal Load}

\begin{figure}[htbp]
    \centering
    \includegraphics[width=0.45\linewidth]{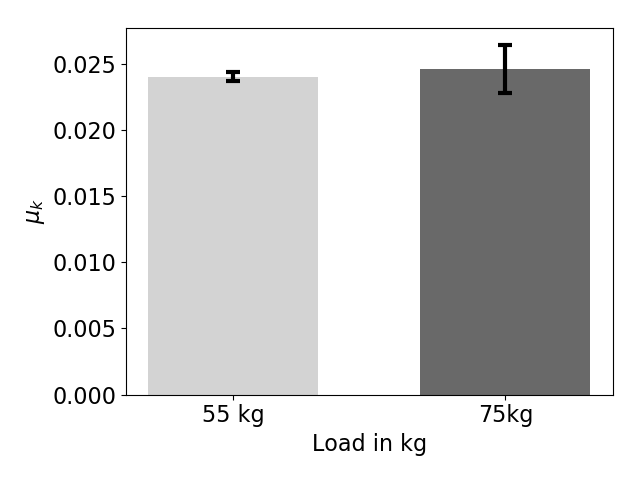}
    \caption{Mean and error bars (2 $\sigma$) for the friction coefficient
    $\mu_k$ for two different normal loads. The sliding velocity was 8 km/h,
    and details about the ambient conditions can be found in Table
    \ref{march15}. We cannot observe any statistically significant difference
    in these conditions.}\label{field_friction_and_load}
\end{figure}

\cite{nachbauer2016friction} conducted full-scale experiments with skis
sliding on snow with a linear tribometer in a general setup similar to ours.
For speeds of 5 and 15 m/s, snow
temperature $T_s=-4^o$C, normal load and $F_N=146$N they measured the friction
coefficient to be independent of the normal force. Similarly, \cite{buhl2001}
found no detectable difference between different loads for temperature above
$-6^o$C. By contrast, \cite{baurle2006sliding} performed experiments with
different load in three different (ambient) temperature regimes. At low
temperatures ($T_{amb} = -10^o$C), no dependency of the friction coefficient on the normal load was found. At intermediate
temperatures ($T_{amb} \approx -5^o$C), the coefficient of friction increased
slightly with load. At temperatures close to the melting point ($-0.5^o$C $<
T_{amb} < 0.5^o$C) the friction coefficient decreased with increasing load.
From our conducted field experiments there was only one valid dataset (March
15) that could be used for investigation of the dependency between friction and
load. The conditions during this test sequence can be found in Table
\ref{march15}. In our case, as shown in Fig. \ref{field_friction_and_load},
the coefficient of friction is found to be independent w.r.t. load. This is
in agreement with \cite{nachbauer2016friction, buhl2001}. 

\begin{table}[htbp]
    \begin{center}
\begin{tabular}{ccccc}
\hline
Time & Ambient temp. ($^o$C) & Snow temp. $(^o)$C & Wind (m/s)& Weather\\
 13:00-15:00 & $-2$ to $-1$ & $-4$ to $-3$ & 2-3 m/s &  Clouded/sun \\
\hline
~\\
\end{tabular}%
    \caption{Weather conditions during the experiments on March 15.}\label{march15}
    \end{center}
\end{table}

\subsection{Measurements of under-ski temperature}

Having presented the results regarding friction, we now turn to the temperature
measurements provided by the thermistors placed along the bottom of the ski.

\subsubsection{Effect of Surface Roughness and the use of cold contra wet Grind}

A highly relevant question from both a technical and scientific point of view
is whether the system is able to detect a difference in temperature at the
ski-snow interface comparing the cold to the wet grind. The friction
coefficient may be higher or lower for either depending on the circumstances,
but the cold grind (due to its finer roughness which results in a larger area
of contact) should yield a higher temperature increase at the ski-snow
interface than the wet grind under the same conditions. 

The first dataset that is studied is from March 14. Two sequences with cold
grinded skis are compared with two sequences with wet grinded skis. All
sequences have the same load, while the velocity was changed between some of
the runs. The details of the parameters used for the different runs are
summarized in Table \ref{temp1403v8_table} (low velocity) and Table
\ref{temp1403v14_table} (moderate velocity).

\begin{table}[htbp]
    \begin{center}
\begin{tabular}{lcc}
\hline
   & Cold grind & Wet grind \\ \hline
Snow temp. ($^o$C) & -5.8 & -4.5 \\
Ambient temp. ($^o$C) & 0 & 2 \\
Time interval of sequence & 10:25-10:37 & 12:15-12:25 \\
\hline
~\\
\end{tabular}%
\end{center}
\caption{Temperatures and time for the test sequences using cold and wet grind
    at low speed conducted 14 March. All test are with a mass of the sledge $m=75$ kg and a
    towing velocity $v=8$ km/h.}\label{temp1403v8_table}
\end{table}

\begin{table}[htbp]
    \begin{center}
\begin{tabular}{lcc}
\hline
   & Cold grind & Wet grind \\ \hline
Snow temp. ($^o$C) & -5.2 & -3.5 \\
Ambient temp. ($^o$C) & 1.2 & 3.1 \\
Time interval of sequence & 10:40-10:50 & 12:30-12:40 \\
\hline
~\\
\end{tabular}%
\end{center}
   \caption{Temperatures and time for test sequences using cold and wet grind
    at high speed conducted 14 March. All test are with a mass of the sledge $m=75$ kg and a
    towing velocity $v=14$ km/h.}\label{temp1403v14_table}
\end{table}

\begin{figure}[htbp]
   \centering
   \includegraphics[width=0.45\linewidth]{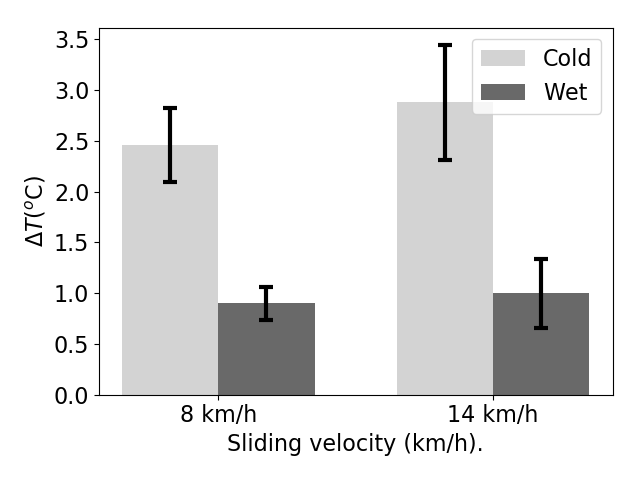}
   \caption{Mean and error bars ($2 \sigma$) of the maximum $\Delta T$ obtained
    from four consecutive runs (for wet and cold grinded skis for two different
    velocities) on March 14th.}\label{1403_temp_T2}
\end{figure}

The results from these measurements are shown in Fig. \ref{1403_temp_T2}. The
ski pair with cold grind yields a higher $\Delta T_{max}$ at the location of
the thermistor $T2$ for both low and moderate velocity. These measurements are
in agreement with theory, i.e. the cold grinded ski base yields higher
frictional heating due its larger area of contact with snow compared to a wet
grinded ski.

A similar analysis can be done for the test sequences conducted on
March 27. Due to time constraints, only one comparable set of test sequences
was obtained, which details can be seen in Table \ref{temp2703v8_table}. The
ambient and snow temperature were lower during this experiment, and the results
from $T1$ (closets to ski tip) are included.

\begin{table}[htbp]
    \begin{center}
\begin{tabular}{lcc}
\hline
   & Cold grind & Wet grind \\ \hline
Snow temp. ($^o$C) & -7 & -5 \\
Ambient temp. ($^o$C) & -1 & -0.1 \\
Time interval of sequence & 10:50-11:05 & 12:15-12:30 \\
\hline
~\\
\end{tabular}%
   \caption{Temperatures and time corresponding to test runs for cold and wet
    grind conducted 27 March. Both with $m=75$ kg and $v=8$
    km/h.}\label{temp2703v8_table}
\end{center}
\end{table}

\begin{figure}[htbp]
    \centering
    \includegraphics[width=0.45\linewidth]{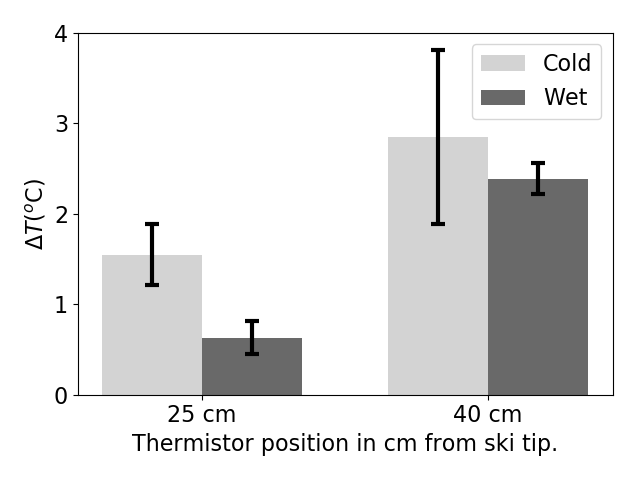}
    \caption{Mean and standard deviation of the four maximum $\Delta T$
    obtained from the four runs (for each ski) on March 27th. The thermistor at 140 cm on the
    cold grinded ski is broken, hence no temperature was
    measured.}\label{temp2703_bars}
\end{figure}

The results presented in Fig. \ref{temp2703_bars} show similar trends to the
result obtained on March 14, but they are statistically less significant with
respect to the $2 \sigma$ confidence intervals. 

\subsubsection{Effect of Velocity}\label{tempandvel}

An increase in temperature at the ski-snow interface, corresponding to the
appearance of a thin lubrication layer of water, is expected to occur as soon as
reasonable cross-country skiing velocities are achieved. Hence, we investigate the
correlation between velocity and temperature. More specifically, if one
considers in first approximation that all the energy lost due to ski-snow
friction is turned into heat, and one neglects both heat transfer and phase
changes, then the heat flux provided to the snow can be calculated as:

\begin{equation}\label{heatgeneration}
    q = \frac{\mu_k F_N v}{A},
\end{equation}

\noindent where $A$ is the area of the slider and $F_N=mg$ is the normal load.
This heat flux is at the origin of both the phase change from snow to liquid
water phase, and temperature increase of the interfacial water layer, which
together consitute a complex, non-linear system \citep{baurle2006sliding, baurle2007sliding}.

\begin{figure}
    \begin{center}
   \includegraphics[width=0.45\linewidth]{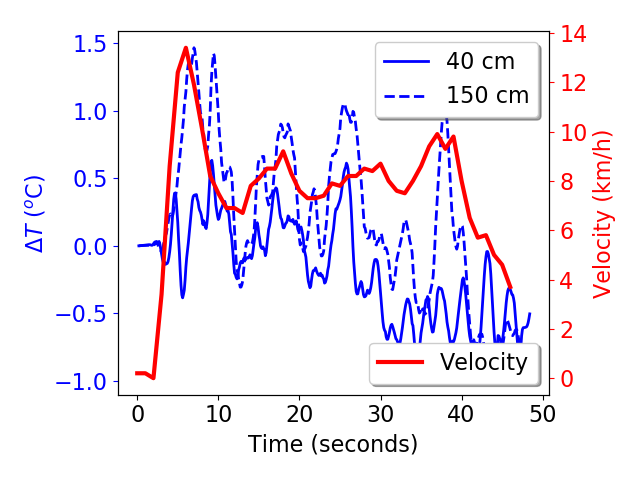}
   \includegraphics[width=0.45\linewidth]{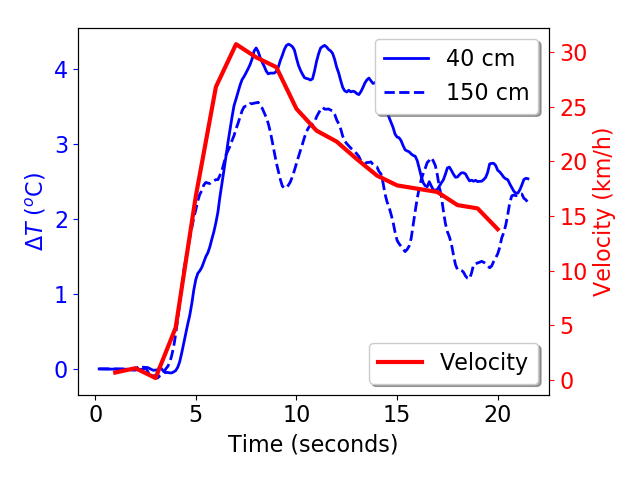}
   \caption{Velocity and $\Delta T$ for a run with maximum velocity
    \(v_{max}\approx 14\) km/h (left) and \(v_{max}\approx 30\) km/h (right). The two sensors which experience the highest
    increase in temperature are shown. The left axis shows temperature and the
    right axis shows velocity.}\label{figtemperature}
        \end{center}
\end{figure}

From Eq. (\ref{heatgeneration}) one can observe that the heat generation is
proportional to $v$. Fig. \ref{figtemperature} confirms the correlation between
velocity and temperature. For lower velocities the temperature increase is
smaller. Notice that for the highest velocity it is the sensor in front of the
ski binding that shows the highest $\Delta T$, while for the lower velocity it is
the sensor on the rear side of the binding that has the higher temperature increase. It is worth mentioning that the correlation between velocitiy and heat generation is presumably not linear. A higher heat flux (due to higher velocity) can lead to increased water film thickness, which in turn can lead to lower friction and thus lower increase in temperature \cite{baurle2006sliding}.

\section{Discussion}
\subsection{Measurements concerning coefficient of friction}
When investigating the coefficient of friction's dependence of environmental conditions like air and snow temperature, one would ideally use the latter as the main parameter. But since the snow surface temperature is very difficult to measure due to solar and infra-red radiation effects on sensors, we used the air temperature in this investigation, although it is the temperature in the slider-snow interface that is of primary
interest.

Our experiments concerning coefficient of friction and the effect of surface roughness of the skis showed no difference between using cold and wet grind preparation. One obvious shortcoming for these results were the lack of valid datasets; we had only three valid sets to analyze this dependency. It would also been interseting to investigate this dependency during conditions with lower ambient temperature. Furthermore, since friction is generally low, it is difficult to perform field tests to show its variations w.r.t. roughness. There may be other properties that influence the friction more, e.g. change in ambient temperature, snow temperature and track conditions during testing, making it hard to differentiate the difference in the friction coefficient $\mu_k$ due to different grind.

Another study that would have benefitted of more data was the exploration of the velocity's effect on friction. In our experiments there was only one
occasion where the temperature was low enough to investigate the decrease in
friction with increasing velocity. Unfortunately, some of the load cell data
from this trip were useless due to disturbances from the sled over-running the rope, hence no comparison from this field trip could be performed. In the remaining field
experiments the snow temperature was close to zero. Thus we are only be able to
investigate the friction coefficient dependence on velocity in the warmer
temperature regime. For each date, all runs (both high and low velocity) were
conducted within 30 minutes, hence the snow conditions for the high and low
velocity runs of each series should be similar.

Our results found the coefficient of friction to be independent w.r.t. load. According to
\cite{baurle2006sliding}, one might expect a difference in either direction,
since our experiments were conducted in an temperature higher than his
intermediate regime and lower than his close-to-zero regime. However, our
experimental dataset remains quite small. In order to provide a more thorough
and accurate analysis of this dependency, further data collection would be
necessary.

\subsection{Measurements concerning under-ski temperature}
The biggest concern regarding the analysis of effect of surface roughness and the use of cold contra wet grind is the lack of comparable datasets. Out of the seven
conducted field experiments there were only three occasions where both wet and
cold grind where tested. One of these three experiments was conducted at
Holmenkollen on the 10th of April in warm weather ($T > 0^o$C), hence the
temperature evolution beneath the ski is expected to be marginal (if not
absent), due to the presence of melt water on the tracks. In all cases, the
three thermistors placed closest to the rear end of the skis showed little or
no change in temperature. These thermistors had very little or no contact with
the snow surface, hence their measurements will be omitted from this
presentation. Moreover, some of the thermistors (most likely due to their vertical position
in the ski base) showed a temperature close to zero when the sled was at rest,
which may indicate that they were too prominent and dented inside the snow. In
addition, there was an unfortunate period of dead time between the runs with
cold grinded skis and wet grinded skis since we had only one experimental setup
and we needed to change the skis between wet and cold grind runs, and during
this period the ambient temperature increased. As a consequence, we chose to focus
on the thermistor which was most exposed and closest to the ski-snow interface,
namely $T2$, the thermistor placed 40 cm from the ski tip.

The main source of error
regarding these results is the non-negligible amount of time elapsed between
respective sequences, during which the air and snow temperatures, as well as
the track and weather conditions, had changed. This affects the measurements,
but given the information at hand it is hard to state to which extent.

Concerning the effect of velocity, the runs conducted only have one acceleration phase which makes it difficult to
detect at which velocity a sharp increase in temperature occurs. Given a flat
and straight test track that is long enough, one could perform runs with
several acceleration phases, i.e. stepwise acceleration at given time intervals. Such
an approach could help reveal the relation between velocity and temperature in
the ski-snow interface, i.e. characterize the temperature as a function of
velocity, and allow to find the exact critical velocity when transition happens.

\subsection{Recommendations for future work}
The test period that lasted from February 28th to April 10th revealed several aspects 
that could be improved in further deployments and uses of the experimental setup:

\begin{itemize}
    \item A long, straight and flat test track is beneficial when conducting
        experiments with the ski-tribometer. Such a track can be difficult to
        obtain, hence an appropriate device for measuring the elevation profile
        of the chosen track should be used to allow compensating for elevation gradient
        if these are present. With such information one could
        estimate the contribution from gravity due to the inclination of the
        slope which would provide much higher accuracy when calculating the
        friction coefficient $\mu_k$ in the field.
    
    \item Since we observed that the friction coefficient $\mu_k$ in cross-country skiing is dependent
        on the velocity it would be useful to perform even more accurate
        velocity measurements. This could for example be performed using a GPS with higher
        accuracy and sampling rate. A rotary encoder attached to the winch
        would also be able to provide more precise velocity measurements. Such
        a device would give a more detailed picture of how the winch operates
        given the motor power input from the user. This feature could provide
        useful information regarding how the winch and the motor would respond
        to varying friction during a test run, and allow to build a closed-loop
        control similarly to what was used in a different context by \cite{rabault2019experiments}.
        Velocity data with higher accuracy would be important with respect to
        comparison to the load cell data, i.e., to ensure that the velocity
        truly is constant over the selected measurement section. More
        importantly, if a phase transition was to occur, the friction would
        drop and an increase in velocity could be detected.
    
    \item  The temperature measurements showed promising results, hence more
        thermistors should be placed under the ski. This would give an even
        more detailed picture of the temperature profile. Attaching sensors
        under both skis should be done for redundancy. This would make the
        setup less vulnerable if some of the thermistors were to break. It is
        important that all sensors have identical setup and vertical
        positioning, thus great care should be taken when placing the
        thermistors in the holes. Therefore, a setup which would allow a more
        precise adjustment of the thermistors should be developed.
    
    \item A more inelastic static rope should be used when pulling the slider
        to avoid building up tension during a run.
    
    \item Detailed measurements of snow structure and weather parameters were
        not in the scope of this work. This is something that should be
        considered if someone is to continue working with a similar setup. Such
        characteristics are crucial if one is to really understand the
        measured values from ski friction $\mu_k$ and the temperature
        evolution at the ski-snow interface \cite{fauve2005influence}.
\end{itemize}

In addition, further in-depth modifications of the setup could be performed at
the expense of more work. For example, one could attach a rope both in front and
behind of the ski sledge, and use two winches (one ahead and one behind the skis
sledge) to avoid 'slack' and 'catapulting' effects. In this case, one should mount
two force gauges on the skis sledge (on at each attachment point of the ropes) to
be able to measure the net force applied on the sledge, and enforce careful synchronization
of both winches through wireless communications.

Furthermore, such a setup would allow to tow several sledges in a row and, therefore,
to measure the friction coefficient of several pairs of ski at once. This may be the
best methodology for actually performing fine-grained comparison of friction between
different pairs of skis, and could be of interest for competition situations where many
pairs of skis must be tested in a short amount of time.

\section{Conclusions}

In the present work, we outlined the design of an outdoor tribometer allowing
to characterize both ski friction on snow and interfacial ski-snow temperatures
in a variety of conditions. The outdoor tribometer allows snow friction
measurements in the field under realistic loads and velocities with respect to
cross-country skiing, which is critical given the recent shift in regulation
regarding banning of fluoro-based gliders. Measured friction coefficients and
temperature profiles under a gliding skating ski are in agreement with reported
values from similar setups. The main findings obtained when investigating the
influence of environment and ski preparation parameters on the coefficient of friction and the
temperature in the ski-snow interface are listed below:

\begin{itemize}
    \item The coefficient of friction on snow decreases for increasing
        temperatures, when the outside temperature is low enough ($T_{amb} \leq
        0^o$C). This is in agreement with several previous studies
        \cite{bow39, klein1947snow, ermakov1984coefficient, buhl2001,
        baurle2006sliding, nachbauer2016friction}. The temperature change increases at
        the ski-snow interface with decreasing temperature due to
        solid-to-solid interactions which results in higher frictional heating
        and, ultimately, the formation of a thin water layer.
    
    \item The coefficient of friction increases with velocity for snow
        temperatures near or above zero ($T_{amb} > -1^o$C, $8 \leq v \leq 27$
        km/h ). Since a water film is expected to already be present at
        these temperatures, further frictional heating generates too much water
        and induces capillary drag.  This is also consistent with the results
        presented by several other groups \cite{kuroiwa1977kinetic,
        jones1994friction, baurle2006sliding}. The temperature change at the ski-snow
        interface increases with increasing velocity due to a higher heat
        generation.
\end{itemize}

Finally, we want to stress that skiing and friction reduction still probably have a lot
to gain from scientific exploration and optimization of the parameter spaces. Probably one
of the challenges for systematic scientific work is the difficulty in enforcing reproducibility
in the preparation of skis, including both the microstructure of the friction area, and 
the exact composition and thickness of the chemical coating applied.

\section{Declarations}

Funding from the DOFI Petromaks2 project (grant number 280625 funded by the Research
Council of Norway) is gratefully acknowledged. We thank The Norwegian Olympic Sports Center, for providing skis used in this study and grinding them, as well as for
several interesting discussions in the early steps of this work.
The authors declare no conflicts of interest in this study.
All the code for the experimental setup (both winch control system, and measurements
electronics on the sledge) is released on github (link: [to be released upon
publication in the peer-reviewed literature]). Raw data acquired during the measurement
campaigns and processing scripts are available upon request to the corresponding
author.

\bibliographystyle{ametsoc2014}
\bibliography{bibliography}

\end{document}